\documentclass[aps,floatfix,nofootinbib,amsmath,amssymb,showpacs]{revtex4}

\usepackage[dvips]{graphicx,color}
\usepackage{mathrsfs}
\newcommand{\be}{\begin{equation}}
\newcommand{\ee}{\end{equation}}

\def\lsim{\mathrel{\raise.3ex\hbox{$<$\kern-.75em\lower1ex\hbox{$\sim$}}}}
\def\gsim{\mathrel{\raise.3ex\hbox{$>$\kern-.75em\lower1ex\hbox{$\sim$}}}}

\begin{document} 

\title{Pulsars as the Sources of High Energy Cosmic Ray Positrons} 

\author{Dan Hooper}
\affiliation{Theoretical Astrophysics, 
Fermi National Accelerator Laboratory, Batavia, USA}
\affiliation{Department of Astronomy and Astrophysics, 
The University of Chicago, USA}

\author{Pasquale Blasi}
\affiliation{Theoretical Astrophysics, 
Fermi National Accelerator Laboratory, Batavia, USA}
\affiliation{INAF-Osservatorio Astrofisico di Arcetri, Firenze, Italy}
\affiliation{INFN-Laboratori Nazionali del Gran Sasso, Assergi,
  L'Aquila, Italy}

\author{Pasquale Dario Serpico}
\affiliation{Physics Department, Theory Division,
CERN, CH-1211 Geneva 23, Switzerland}
\affiliation{Theoretical Astrophysics, 
Fermi National Accelerator Laboratory, Batavia, USA}

\begin{abstract}
Recent results from the PAMELA satellite indicate the
presence of a large flux of positrons (relative to electrons) in the
cosmic ray spectrum between approximately 10 and 100 GeV. As
annihilating dark matter particles in many models are predicted to
contribute to the cosmic ray positron spectrum in this energy range, a
great deal of interest has resulted from this observation. Here, we
consider pulsars (rapidly spinning, magnetized neutron stars) as an
alternative source of this signal. After calculating the contribution
to the cosmic ray positron and electron spectra from pulsars, we find
that the spectrum observed by PAMELA could plausibly originate from
such sources. In particular, a significant contribution is expected
from the sum of all mature pulsars throughout the Milky Way, as well
as from the most nearby mature pulsars (such as Geminga and
B0656+14). The signal from nearby pulsars is expected to generate a
small but significant dipole anisotropy in the cosmic ray electron
spectrum, potentially providing a method by which the Fermi gamma-ray
space telescope would be capable of
discriminating between the pulsar and dark matter origins of the
observed high energy positrons.  
\end{abstract}
\pacs{98.70.Sa;97.60.Gb;95.35.+d;98.70.Sa\hfill FERMILAB-PUB-08-429-A, CERN-PH-TH/2008-228}
\maketitle

\section{Introduction}

Dark matter particles annihilating in the Galactic Halo are predicted
to generate a number of potentially observable products, including
gamma-rays, electrons, positrons, protons and antiprotons. In contrast
to gamma-rays, which travel along straight lines, charged particles
move under the influence of the Galactic Magnetic Field, diffusing and
losing energy, resulting in a diffuse spectrum at Earth. By studying
the cosmic ray antimatter spectra, balloon or satellite-based
experiments hope to identify signatures of dark matter.  

The PAMELA satellite, which began its three-year mission in June of 2006, is
designed to measure the spectra of cosmic ray positrons up to 270 GeV
and electrons up to 2 TeV, each with unprecedented
precision~\cite{pamela}. Recent results show that
the ratio of positrons to electrons plus positrons (the positron
fraction) in the cosmic ray spectrum appears to stop decreasing
and begins to climb quite rapidly between 10 GeV and 
100 GeV~\cite{Adriani:2008zr}.  A similar trend was in fact also indicated 
by earlier experiments, including HEAT~\cite{heat} and 
AMS-01~\cite{ams01}, although with lesser statistical
significance.

The behavior in the positron fraction observed by PAMELA is very
different from that predicted for secondary positrons produced in the
collisions of cosmic ray nuclides with the interstellar
medium. Barring systematics (e.g. problems in the positron/proton
separation at high rigidity), the data appear to indicate the
existence of additional, primary sources of high energy 
positrons, such as dark matter annihilations taking place in the halo
of the Milky
Way~\cite{Bergstrom:2008gr,Cirelli:2008jk,Barger:2008su,Cholis:2008hb,Cirelli:2008pk,ArkaniHamed:2008qn}. 
It  
should be noted, however, that while the observed spectral shape can
be easily accommodated with annihilating dark matter, the
normalization of this contribution to the PAMELA data requires a
somewhat large annihilation rate. Such a large rate could, in
principle, result from the dark matter possessing an annihilation
cross section in excess of the value predicted for a simple $s$-wave
thermal relic ($\sigma v \sim 3\times 10^{-26}$ cm$^3$/sec), for example due to
the Sommerfeld
effect~\cite{Cirelli:2008jk,ArkaniHamed:2008qn,sommerfeld} or in the
case that the 
WIMP is of non-thermal origin~\cite{nonthermal}. Alternatively, a
large annihilation rate could be generated as the result of
significant inhomogeneities in the spatial distribution of dark
matter, such as the presence of a large, nearby dark matter
subhalo~\cite{hoopertaylorsilk}. In either case, such scenarios are
somewhat constrained by observations of gamma-rays~\cite{baltz},
antiprotons~\cite{Cirelli:2008pk,nonthermal} and synchrotron
emission~\cite{synchrotron,nonthermal}. 

The challenges involved in explaining the PAMELA signal with
annihilating dark matter lead us to consider a less exotic and purely
astrophysical explanation for the observed positron
flux. Energetic electron-positron pairs can be in fact produced in astrophysical
sources, the leading candidate sites being pulsars -- rapidly
spinning, magnetized neutron stars, which emit pulsed electromagnetic
radiation, as observed from Earth. 

In this paper, we explore the possibility that the positron fraction
reported by PAMELA may be generated by mature
pulsars. Gamma-ray pulsars are predicted to produce energetic
electron-positron pairs with a harder spectrum than that from
secondary cosmic-ray induced origin, leading to the possibility that
such sources may dominate the cosmic ray positron spectrum at high
energies. We calculate the spectrum of such particles from known local
pulsars (Geminga and B0656+14), and from the sum of all pulsars
distributed throughout the Milky Way. As found in earlier
studies~\cite{aharonian}, we find that both local pulsars and the sum 
of pulsars distributed throughout the Milky Way can contribute
significantly to the observed spectrum. At 10 GeV, we estimate that
on average only $\sim$20\% of the cosmic ray positrons originate
from pulsars within 500 parsecs from the Solar System. If gamma-ray
pulsars are formed at a rate of $\sim$4 per century in the Milky Way,
we find that the observed flux of $\sim$10-20 GeV positrons could be
plausibly generated in such objects. Similar conclusions were derived
in \cite{zhangcheng,grimani}. Above $\sim$50 GeV,
however, the positron spectrum is likely to be dominated by a single
or small number of nearby pulsars. If the high energy
electron-positron spectrum is dominated by a single nearby source, it
opens the possibility of detecting a dipole anisotropy in their
angular distribution (see also~\cite{middle}). We find that such a
feature could potentially be 
detected by the Fermi gamma-ray space telescope (formerly known as
GLAST)~\cite{fermi}, thus enabling a powerful test to discriminate
between the pulsar and dark matter origins of the observed cosmic ray
positron excess.  

The remainder of this article is structured as follows: In
Sec.~\ref{astrosource}, we review the known properties of pulsars and
consider them as sources of high energy electron-positron pairs. In
Sec.~\ref{nearby}, we consider the nearby pulsars Geminga and B0656+14
and discuss their potential contributions to the cosmic ray positron
spectrum.  In Sec.~\ref{anisotropy}, we calculate the expected dipole
anisotropy from nearby pulsars and compare this to the sensitivity of
the Fermi gamma-ray space telescope. We summarize and draw our
conclusions in Sec.~\ref{conclusions}.

\section{Pulsars as sources of electron-positron pairs}\label{astrosource}
In both models of polar gap \cite{ruderman75,arons83} and outer gap
\cite{chenghoruderman86}, electrons can be accelerated in different
regions of the pulsar magnetosphere and induce an electromagnetic
cascade through the emission of curvature radiation, which in turn
results in production of photons which are above threshold for pair
production in the strong pulsar magnetic field. This process results
in lower energy electrons and positrons that can escape the
magnetosphere either through the open field lines \cite{chietal96} or
after joining the pulsar wind \cite{zhangcheng}. In this second case,
the electrons and positrons lose part of their energy adiabatically
because of the expansion of the wind. The energy spectrum injected by
a single pulsar depends on the environmental parameters of the pulsar,
but some attempts to calculate the average spectrum injected by a
population of mature pulsars suggest that the spectrum may be
relatively hard, having a slope of $\sim$$1.5$-$1.6$~\cite{zhangcheng}. 
This spectrum, however, results from a complex interplay of individual
pulsar spectra, of the spatial and age distributions of pulsars in the
Galaxy, and on the assumption that the chief channel for pulsar spin
down is magnetic dipole radiation. Due to the related uncertainties,
variations from this injection spectra cannot be ruled out. Typically, one
concentrates the attention on pulsars of age $\sim$$10^5$ years
because younger pulsars are likely to still be surrounded by their
nebulae, which confine electrons and positrons and thus prevent them
from being liberated into the interstellar medium until later times.   

Still, some energetics considerations can be done with simple analytical
models; this will also help the understanding of arguments developed
in the next 
Section. The rate of energy injection from a single pulsar in the form
of pairs  
is limited by its spin-down power (the rate of energy loss
corresponding to the slowing rate of rotation). Assuming that this is
simply due to the emission of magnetic dipole radiation, the maximum
rate of energy injection can be written as (see e.g.~\cite{pulsarRefs}):

\begin{equation}
\dot E = - \frac{B_s^2 R_s^6 \Omega^4}{6 c^3} \approx 
10^{31} B_{12}^2 R_{10}^6 P^{-4} \, \rm erg \,\, s^{-1},
\end{equation}
where $B_{12}=B_s/10^{12} {\rm G}$ is the magnetic field at the surface of
the star, $R_{10}=R_s/10 {\rm km}$ is the radius of the star and $P$ is the
period of the star in seconds. The period $P$ (gyration frequency
$\Omega$) increases (decreases) with time as a result of the
spin-down, according to 

\begin{equation}
\Omega(t) = \frac{\Omega_0}{\left(1+t/\tau_0 \right)^{1/2}},
\end{equation}
where $\tau_0 = 3 c^3 I/(B_s^2 R_s^6 \Omega_0^2)$, $I=(2/5)M_s R_s^2$
is the moment of inertia of the star with mass $M_s$ and
$\Omega_0=2\pi/P_0$ is the initial spin frequency of the pulsar and
$P_0$ is the initial period. Numerically, this yields:

\begin{equation}
\tau_0 = 7.4 \times 10^7 B_{12}^{-2} \left(\frac{M_s}{1.4
  M_\odot}\right) R_{10}^{-4} P_0^2 \, \rm years. 
\label{eq:tau}
\end{equation}
It follows that the upper limit to the rate of energy deposit in the
form of electron-positron pairs is
\begin{equation}
{\cal L} = I \Omega \dot\Omega = \frac{1}{2}I \Omega_0^2
\frac{1}{\tau_0} \frac{1}{\left(1+\frac{t}{\tau_0} \right)^2}.
\label{eq:spindown}
\end{equation} 
In terms of the total energy injected in a time $t$ after the pulsar birth, 
\begin{equation}
E_{tot}(t)=\frac{1}{2}I \Omega_0^2
\frac{t}{\tau_0}\frac{1}{1+\frac{t}{\tau_0}}=6\times 10^{43} P_0^{-4}
R_{10}^6 B_{12}^2 t_{5} \frac{1}{\left(1+\frac{t}{\tau_0}
  \right)} \, \rm erg,
\label{eq:energy}
\end{equation}
where $t_{5}$ is the time in units of $10^5$ years.  Therefore,
the total energy that a mature pulsar ($t\gg \tau_0$) has injected in the
form of magnetic dipole radiation saturates to 
\begin{equation}
E_{tot}\approx\frac{1}{2}I \Omega_0^2 = 2.2\times 10^{46} \left(
\frac{M_s}{1.4 M_\odot} \right) R_{10}^2  P_0^{-2} \, \rm erg.
\label{eq:energymax}
\end{equation}
In the same assumption of a mature pulsar, we also have that
$\Omega_0\approx \Omega (t/\tau_0)^{1/2}$, where $\Omega$ is the
gyration frequency measured today. For instance, for the Geminga
pulsar ($P=230$ ms) $t\approx 370,000$ years and $\tau\approx
10^4$ years (using $B_{12}=1.6$ and $R_{10}=1.5$), one has
$\Omega_0\approx 166\,s^{-1}$ ($P_0\approx 40$ 
ms). For these values of the parameters, the total energy output of
the pulsar is rather large, $E_{tot}\approx 10^{49}$ erg, which could
easily account for the high energy positron flux.    
It is worth stressing, however, that only a small fraction of this
energy will eventually end up in the form of 
escaping electron-positron pairs, and thus this number should be
treated as an absolute upper limit on the pair luminosity of a
single pulsar.  Qualitatively, the combined effect of a declining
absolute luminosity [Eq.~(\ref{eq:spindown})] and of an increasing
escape probability conspire in singling out  typical  
ages of $\sim$$10^5$ years for the pulsars expected to contribute
maximally to the positron flux. 

To proceed in a more quantitative way towards the calculation of the overall
spectrum from Galactic pulsars, one needs to adopt a model for the
$e^{+}-e^{-}$ 
acceleration and escape probability from a single pulsar with a given
magnetic field, period, etc. and 
then integrate over a Monte Carlo distribution of these typical
parameters in a Galactic Pulsar population. The resulting injection
spectrum we adopt follows from such a calculation in
Ref.~\cite{zhangcheng}:  

\begin{equation}
\frac{dN_e}{dE_e} \approx 8.6 \times 10^{38} \dot{N}_{100} \,
(E_e/{\rm GeV})^{-1.6} \exp{(-E_e/80\,{\rm GeV})}\, \rm{GeV}^{-1} \,
\rm{s}^{-1}, 
\label{eq:inje}
\end{equation}
where $\dot N_{100}$ is the rate of pulsar formation in units of pulsars per
century. This expression corresponds to an {\it average} energy output
in electron-positron pairs of approximately $6\times 10^{46}$ erg per
pulsar, i.e. to efficiency $\alt 1\%$ compared with the upper bound
derived above. In the following, we inject this spectrum according to
the spatial distribution of pulsars given in
Refs.~\cite{zhangcheng,distribution}.

Once electrons and positrons are produced, diffusion in the Galactic
Magnetic Field regulates their motion. Unlike previous approaches to
the problem, mostly based on a simple implementation of the leaky box
model, we calculate the effects of propagation by solving the
transport equation for electrons, including synchrotron and inverse
Compton scattering losses:

\begin{eqnarray}
\frac{\partial}{\partial t}\frac{dn_{e}}{dE_{e}} =
\vec{\bigtriangledown} \cdot \bigg[K(E_{e})  \vec{\bigtriangledown}
  \frac{dn_{e}}{dE_{e}} \bigg] 
+ \frac{\partial}{\partial E_{e}} \bigg[B(E_{e})\frac{dn_{e}}{dE_{e}}
  \bigg] + Q(E_{e},\vec{x}),
\label{dif}
\end{eqnarray}
with a free escape boundary condition at 4 kpc above and below the
Galactic Plane. Here $dn_{e}/dE_{e}$ is the 
number density of electrons/positrons per unit energy, $K(E_{e})$ is
the diffusion coefficient and $B(E_{e})$ is the rate of energy loss. 
We adopt $K(E_e) \equiv K_0 (1+E_e/(3 \, {\rm GeV}))^{\delta}$ with
$K_0 = 3.4 \times 10^{28}$ cm$^2$/s and $\delta=0.6$, and $B(E_e)
=-b E^2_e$ with $b=10^{-16} \rm{GeV}^{-1} \rm{s}^{-1}$. $Q$ corresponds to 
the source term described above.

\begin{figure}[t]
\centering\leavevmode
\includegraphics[width=3.4in,angle=0]{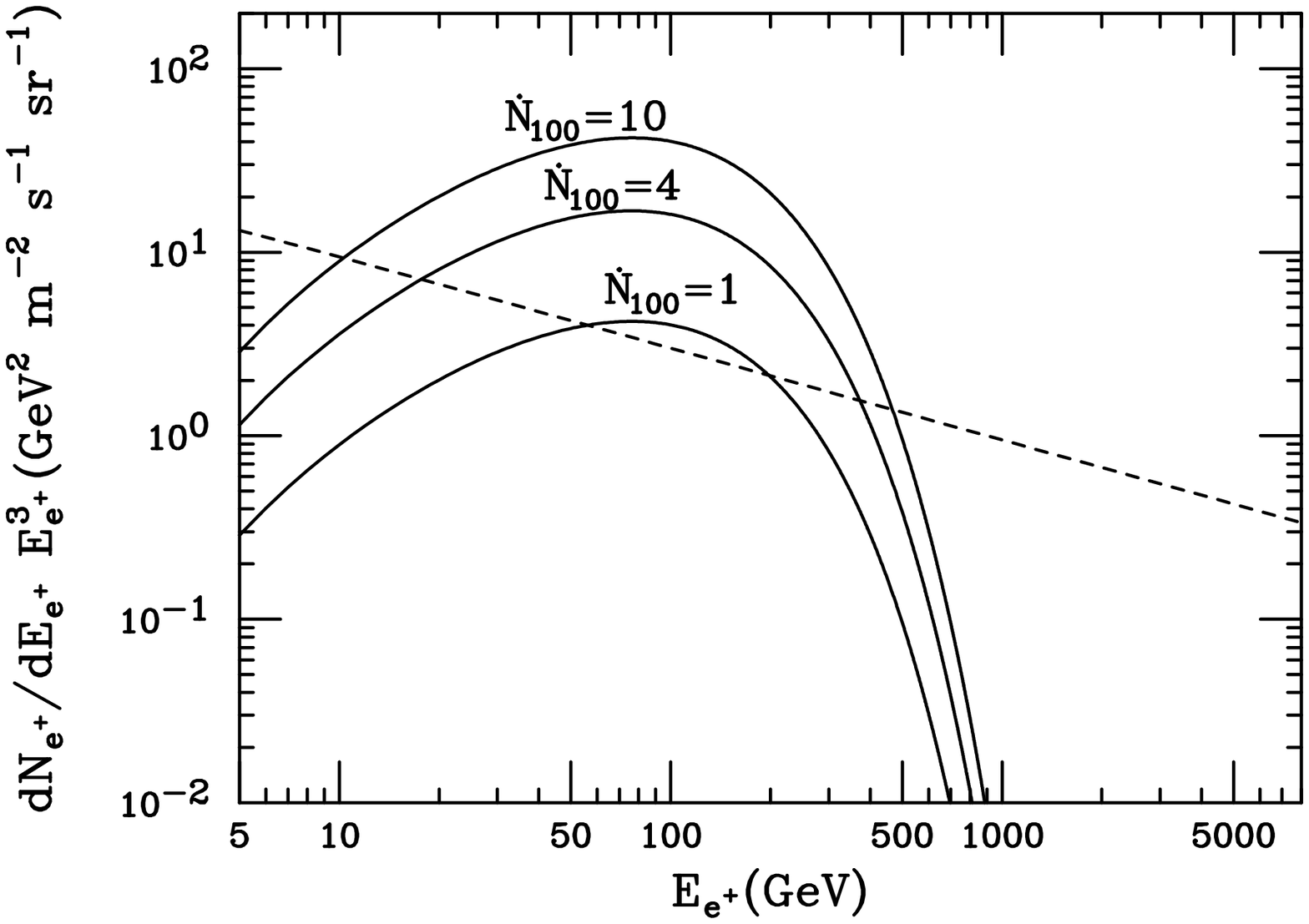}
\hspace{0.1cm}
\includegraphics[width=3.4in,angle=0]{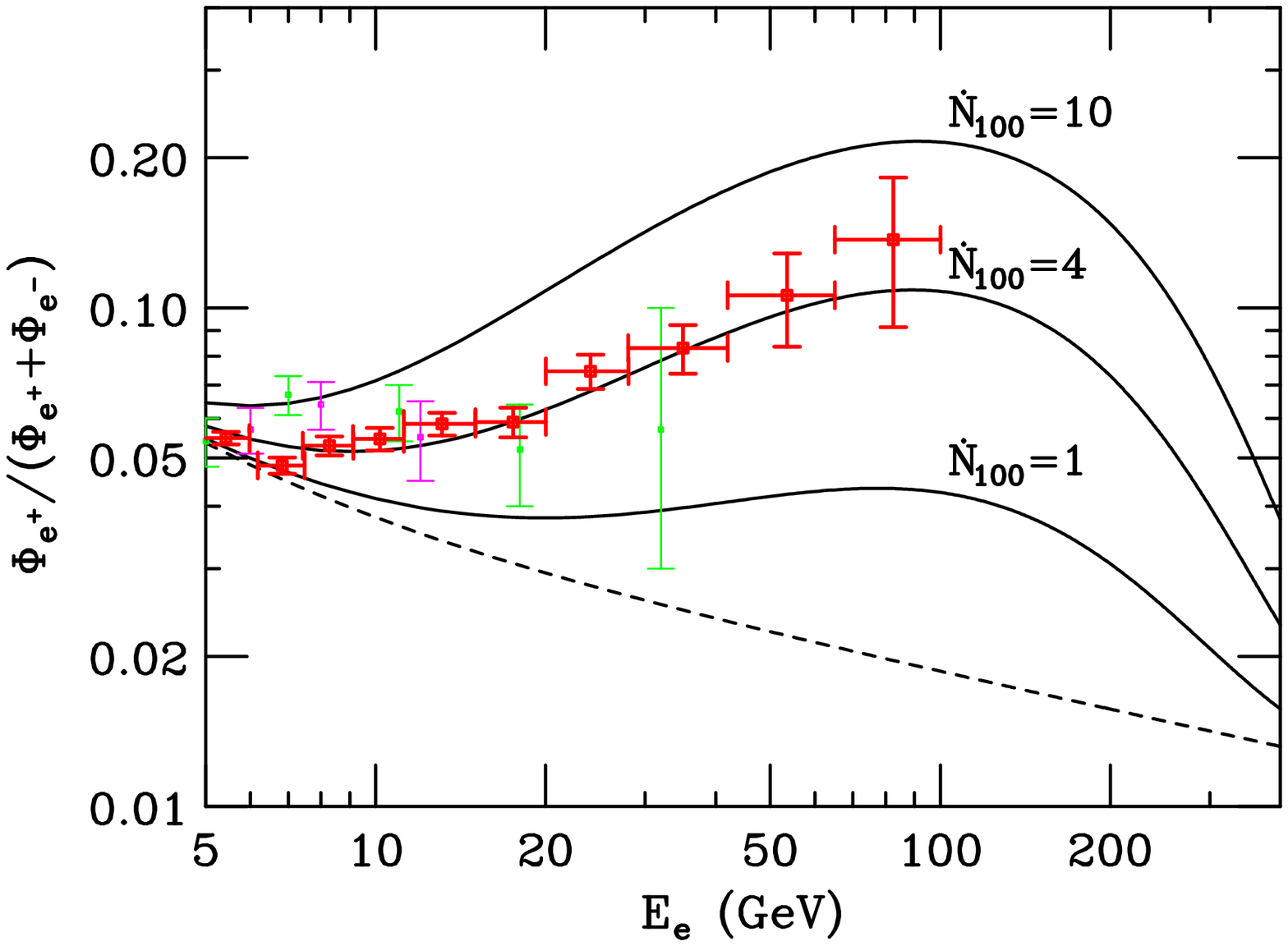}
\includegraphics[width=3.4in,angle=0]{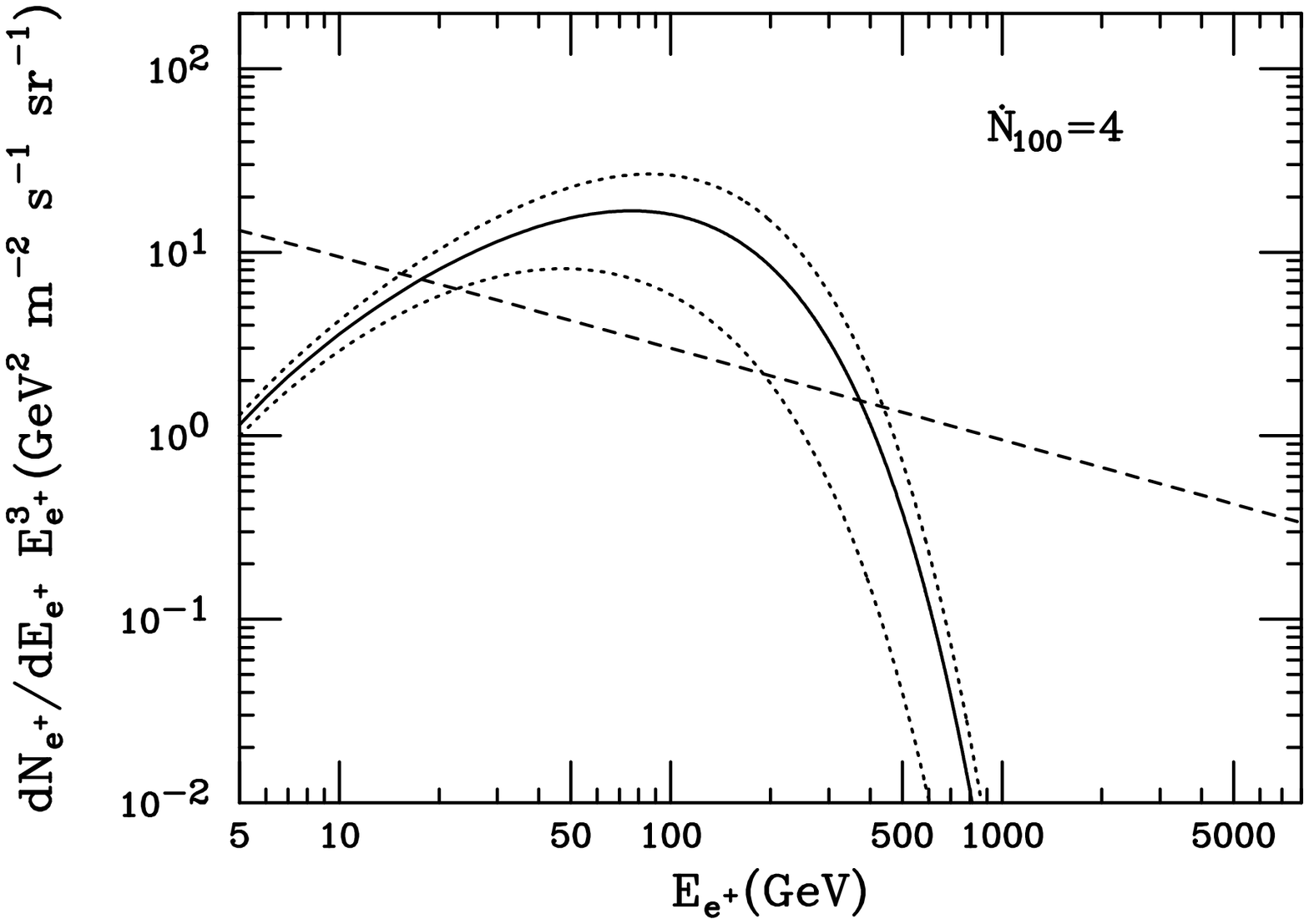}
\hspace{0.1cm}
\includegraphics[width=3.4in,angle=0]{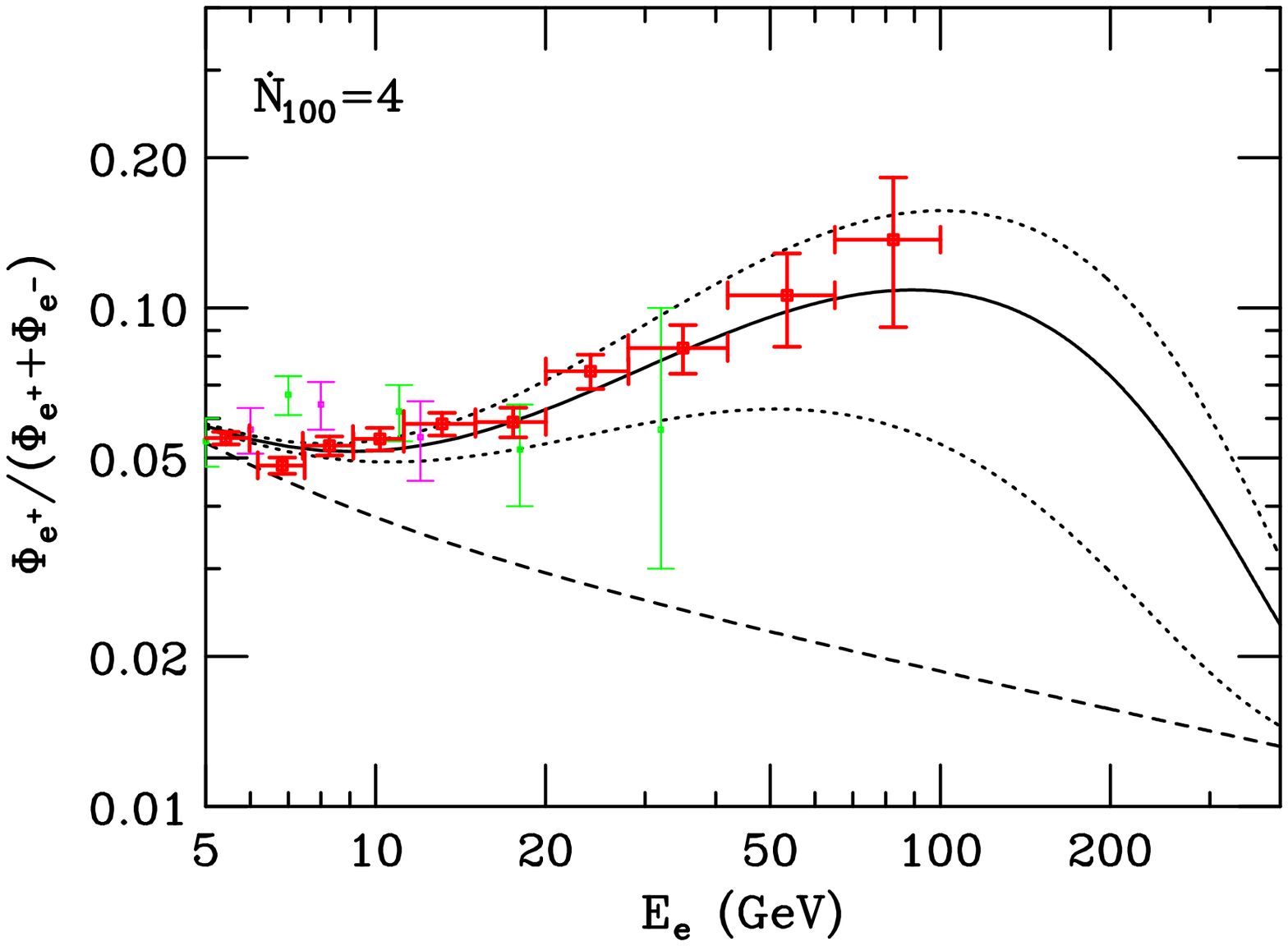}
\caption{The spectrum of cosmic ray positrons (left) and the positron
  fraction (right) resulting from the sum of all pulsars 
  throughout the Milky Way. Also shown as a dashed line is the
  prediction for secondary positrons (and primary and secondary
  electrons in the right frames) as calculated in 
  Ref.~\cite{moskalenkostrong}. In the right frames, the measurements
  of HEAT~\cite{heat} (light green and magenta) and 
  measurements of PAMELA~\cite{Adriani:2008zr} (dark red) are also
  shown. We have used the injected spectrum reported in
  Eq.~(\ref{eq:inje}). In the lower frames, the upper (lower) dotted
  line 
  represents the case in which the injection rate within 500 parsecs of
  the Solar System is doubled (neglected), providing an estimate the
  variance resulting from the small number of nearby  pulsars
  contributing to the spectrum.}  
\label{fig:sum}
\end{figure}

In Fig.~\ref{fig:sum}, we show the spectrum of positrons and the
positron fraction resulting from the sum of all pulsars throughout the
Milky Way. In the upper panels, we show results for different rates of
pulsar birth (one per 10, 25, or 100 years). The dashed line
represents the baseline result neglecting the contribution from pulsars,
including only the positrons produced as secondaries in the hadronic
interaction of cosmic rays. In the right frames, the positron ratio is
obtained considering, besides secondary leptons, also 
the primary electrons accounted as in~\cite{moskalenkostrong}, to ease
the comparison
with previous literature. In the right frames, the measurements of
HEAT~\cite{heat} (light green and magenta) and the measurements of
PAMELA~\cite{Adriani:2008zr} (dark red) are also shown. 

In the lower frames of Fig.~\ref{fig:sum} we show the positron
spectrum and the positron fraction for $\dot N_{100}=4$ if the
injection rate within 500 
parsecs of the Solar System is doubled (upper dotted curve) or neglected
(lower dotted curve). This provides an estimate of the relative
importance of {\it average} nearby sources compared to the
contribution from more distant pulsars. We will discuss this issue
further in Sec. \ref{nearby}. 

Interestingly, the best fit to the HEAT and PAMELA data appears to be
obtained for $\dot N_{100}=4$, namely about one pulsar birth each
$\sim 25$ years. It is worth noting that this number is only slightly
higher than the typical estimates of the galactic core collapse
supernovae rate, from which pulsars are formed. This rate has been
estimated in a variety of ways, including from the scaling of rates in
external galaxies, from the measured gamma-ray flux from galactic
$^{26}$Al, from historical observations of galactic supernovae, and from
empirical upper limits from neutrino observatories~(for a review, see
Ref.~\cite{Raffelt:2007nv}). Also note that since the primary electron
flux is determined from a fit to the absolute flux, which has uncertainties
as large as $\pm 50\%$ around 10 GeV (see the cosmic ray review in
Ref.~\cite{Amsler:2008zz}),  
the best-fit value of $\dot N_{100}$ extracted from the ratio is
affected by at least an error as large. 
Additionally, in principle our numerical results could be modified if
a different 
normalization for the diffusion coefficient were chosen; yet, the constraint
on the confinement time inferred from measurements of the
boron-to-carbon ratio does not leave much freedom in this respect, at
least in the energy region of interest here. The slope of the
diffusion coefficient in energy is also expected not to be a critical
parameter, since the positron ``excess'' has thus far been detected over a
relatively narrow region in energy, where $\delta$ should be virtually
constant. Future observations will determine
whether the actual excess positron spectrum extends to energies beyond
$\sim$100 GeV.  

The cutoff in the positron spectrum derived in our calculations is
solely the result of the corresponding cutoff in the injection
spectrum shown in Eq.~(\ref{eq:inje}).  This cutoff is determined by the
details of the development of the electromagnetic cascade in the pulsar
magnetosphere and, even more importantly, by the distribution of
periods, magnetic fields, and radii of mature pulsars. The exact value
of the cutoff energy should, therefore, not be considered to be a robust 
prediction of the theory, although it represents a good estimate of
the order of magnitude of the cutoff energy. For instance, in
Ref.~\cite{chietal96} it is argued that the typical energy of
electrons and positrons in the cascade associated with the polar gap
is  
$E_e \approx  8 B_{12}^{5/7} P_0^{-17/7} \, \rm GeV\,$
which, for a period of 200 ms, would yield a maximum energy of the
resulting pairs of $\sim 400$ GeV. On the
other hand, the description of Ref.~\cite{zhangcheng} leads to different
maximum energies ($\sim$$80$ GeV) and even different scalings of the
cutoff with the parameters of the pulsar.  It is therefore not very
difficult 
to accommodate moderately higher energy positrons in the pulsar
scenario, if future observations were to show that this is the case. 
Having listed possible caveats to the above results, 
what is remarkable is that, without any fine-tuning of the models, 
typical predictions for both the spectrum and the normalization 
can explain the positron fraction data very well.

At energies lower than $\sim$10 GeV, pulsars are not expected to
contribute any appreciable flux because of the very hard spectrum,
compared to the spectrum of secondary positrons produced in hadronic
interactions of cosmic rays diffusing throughout the Galaxy. These
secondary electrons approximately reproduce the steep spectrum of the
parent nuclei and at low energies dominate the observed positron
fraction. Since the spectrum of positrons from pulsars is important
only at relatively high energies, we have neglected here the role of solar
modulation. 

\section{Nearby Pulsars As A Source of High Energy Cosmic Ray
  Electrons and Positrons}
\label{nearby} 
 
In this section, following earlier studies~\cite{aharonian,middle}, we
re-explore the possibility that an individual or small number of
nearby 
pulsars dominate the cosmic ray positron spectrum within the energy
range studied by PAMELA. As argued in the previous section, in order
to contribute significantly such a pulsar can be neither too young nor
too old. The electrons and positrons from very young pulsars
(i.e. much younger than 100,000 yrs or $t_5 \ll 1$) are expected to be
confined by the surrounding pulsar wind nebula. Furthermore,
electrons/positrons in the PAMELA energy range require $\sim 10^4 \,
(D/100 \, {\rm pc})^2$ years to diffuse a distance, $D$. The
electrons/positrons from very old pulsars, in contrast, will have
diffused over a large volume, diluting their contribution to
insignificant levels. There are at least two pulsars that appear to
satisfy these constraints, Geminga which is estimated to be 157
parsecs from Earth and 370,000 years old, and B0656+14, which is
estimated to be 290 parsecs from Earth and 110,000 years old. 

The spectrum of electron positron pairs at Earth is again calculated
by solving the transport equation, but this time for a single
source. Moreover we consider the case of a bursting source, namely one
in which the duration of the emission is much shorter than the travel
time from the source. For the pulsars discussed above this seems
indeed to be appropriate. For the simple case of a power-law spectrum
of electrons  
and positrons, $Q(E_e) \propto
E_e^{-\alpha}$, injected a time $t$ ago at a distance $D$ from the Solar
System, the spectrum observed at Earth is given by:  

\begin{equation}
\frac{dN_e}{dE_e} = \frac{Q(E_e)}{\pi^{3/2} D^3} (1-b t
E_e)^{\alpha-2}\, \bigg(\frac{D}{D_{\rm dif}}\bigg)^3 \, e^{-(D/D_{\rm 
    dif})^2},  
\end{equation}

where the diffusion distance scale is given by:
\begin{equation}
D_{\rm dif}(E_e,t) \approx 2\sqrt{K(E_e) t \frac{1-(1-E_e/E_{\rm
      max})^{1-\delta}}{(1-\delta) \, E_e/E_{\rm max}}}.
\label{ddifeq}
\end{equation}
As a result of energy losses, the maximum energy that electrons can
reach the observer with is 
approximately given by $E_{\rm max} \approx 1/(b \, t) \approx
3\,t_5^{-1}\,{\rm TeV}\,$. Since energy losses for 
electrons with energy $\sim$$10$-$100$ GeV are not crucial over a time
period of a few hundred thousand years, Eq.~(\ref{ddifeq})
approximately reduces to $D_{\rm dif}(E_e,t)\sim \sqrt{4K(E_e)\,t}$,
corresponding to $\sim$300-500 $\sqrt{t_5}$ parsecs over the energy
range of 10 to 50 GeV.

\begin{figure}[t]
\centering\leavevmode
\includegraphics[width=3.4in,angle=0]{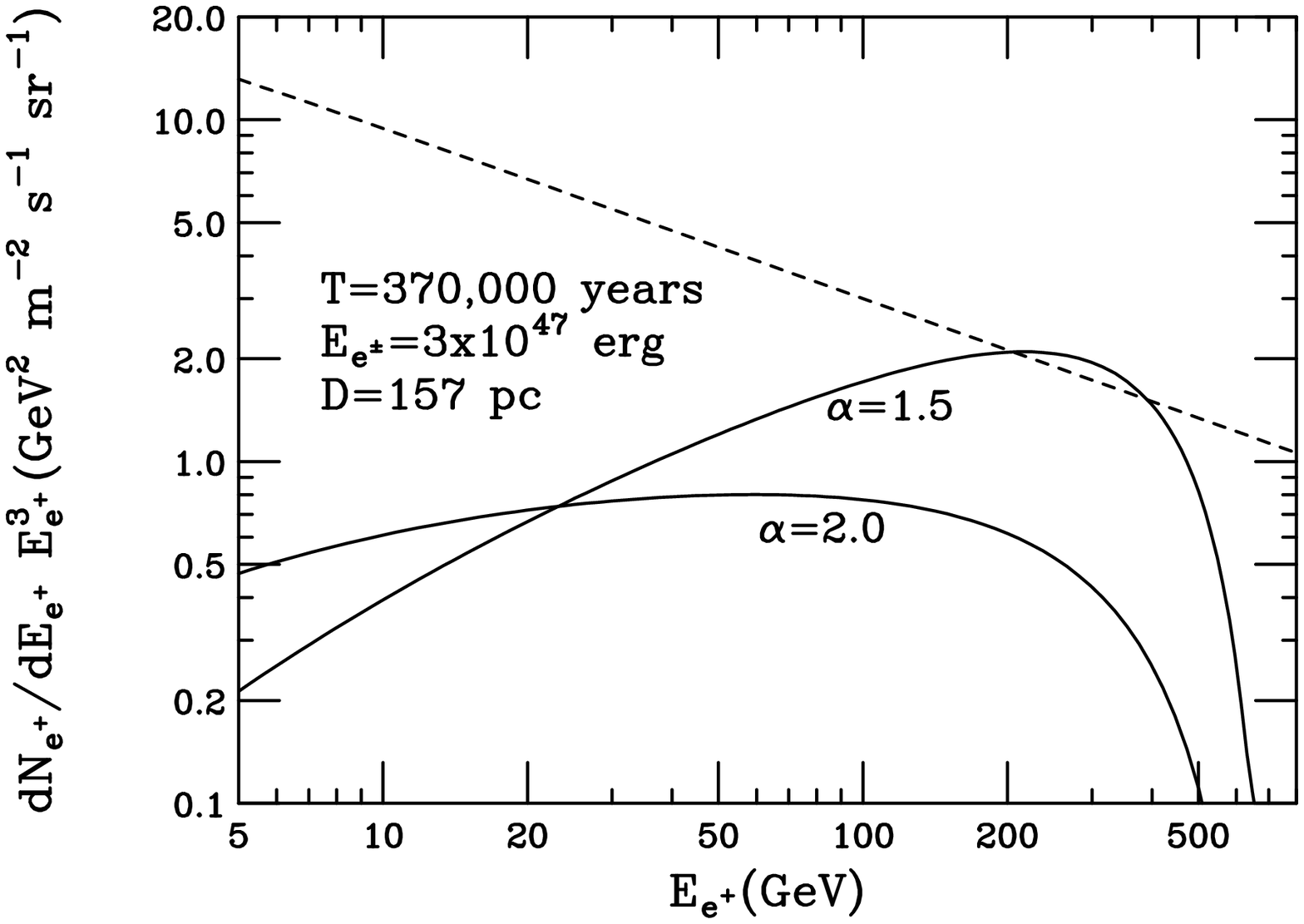}
\hspace{0.1cm}
\includegraphics[width=3.4in,angle=0]{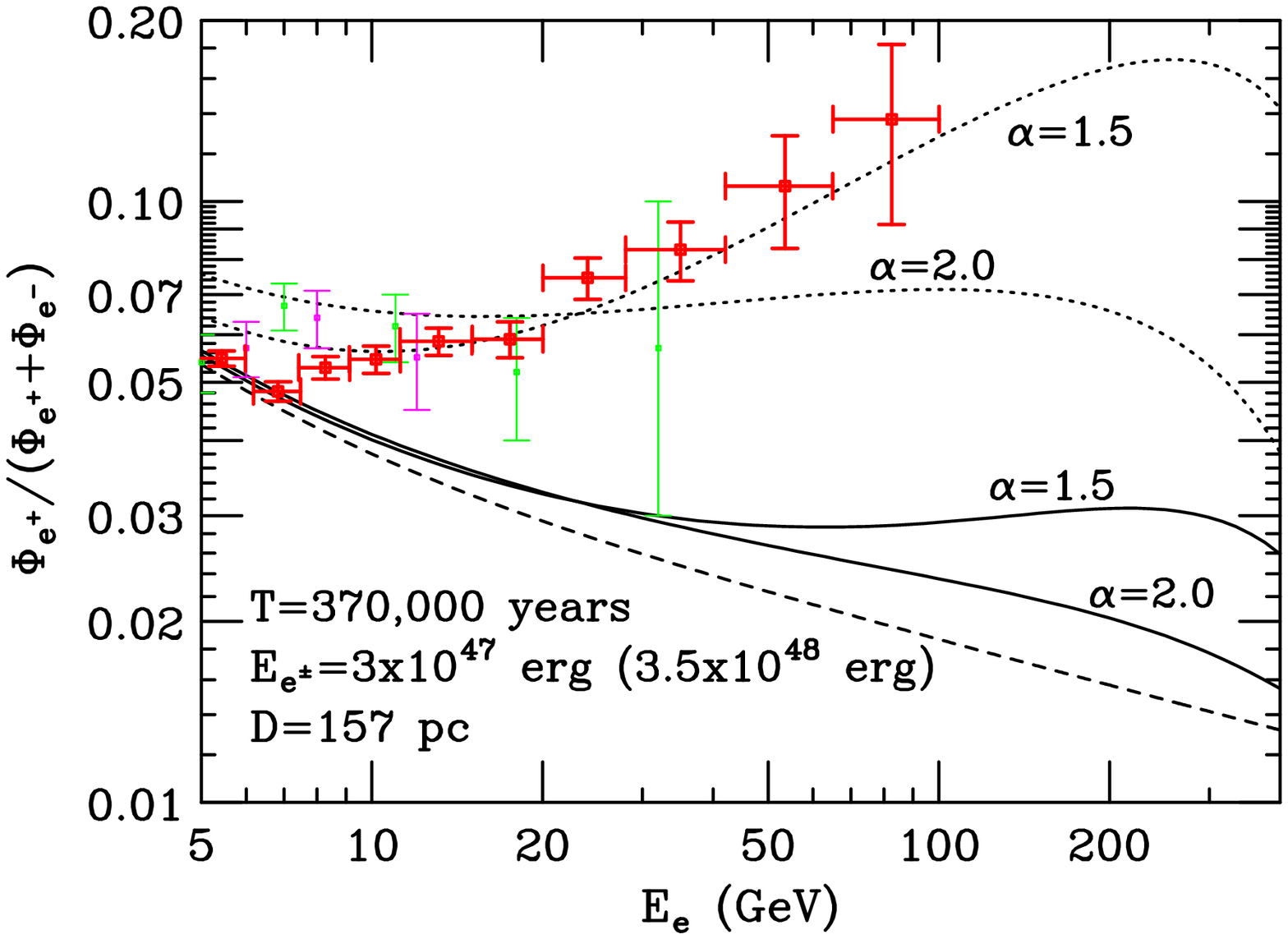}
\caption{The spectrum of positrons (left) and ratio of positrons to
  electrons plus positrons (right) from the pulsar Geminga, with the
  dashed lines 
  as in Fig. 1. In the right frames, the
  measurements of HEAT~\cite{heat} (light green and magenta) and
  measurements of PAMELA~\cite{Adriani:2008zr} (dark red) are
  also shown. Here we have used an injected spectrum such that
  $dN_e/dE_e \propto E^{-\alpha}\, \exp(-E_e/600\,{\rm GeV})$, with
  $\alpha=1.5$ and 2.2.   The solid lines correspond to an energy in
  pairs given by $3.5\times 10^{47}$ erg, while the dotted lines require
  an output of $3\times 10^{48}$ erg.} 
\label{geminga}
\end{figure}

As a first example, we consider the nearby ($D \simeq 157$ pc) Geminga
pulsar, having an estimated age of $370,000$ years~\cite{aharonian,middle}. In
order to take into account the uncertainties in the injection
spectrum, we carry out our calculations by assuming that electrons and
positrons are injected with a spectral index $\alpha=2$ and $1.5$, and
exponentially cutoff above 600 GeV. As a default quantity, we consider
a total energy of $3 \times 10^{47}$ erg injected as electron-positron
pairs, which constitutes a few percent of the total spin down power of
the pulsar.  
Our results for Geminga are shown in Fig.~\ref{geminga}. In the left
panel we plot 
the positron spectrum from Geminga for $\alpha=$1.5 and 2. Again, the
dashed line presents the spectrum from secondary positrons alone. The
right panel shows the positron fraction for the two values of $\alpha$
and for two values of the total energy injected in pairs. The lower of
these values in our default choice, $3\times 10^{47}$ erg (solid
lines), while the higher, dotted lines represent the approximate
energy required to generate the entire flux of excess positrons from
Geminga alone ($3.5\times 10^{48}$ erg). We thus conclude that if
Geminga were to dominate the observed positron fraction at high
energies, it would have to transfer on the order of $\sim$30\% of its
spin-down power into electron-positron pairs. Such a high efficiency
to pairs appears unlikely.  The (probably) subdominant role of Geminga
is not particularly unexpected,  given its relatively old age.

\begin{figure}[t]
\centering\leavevmode
\includegraphics[width=3.4in,angle=0]{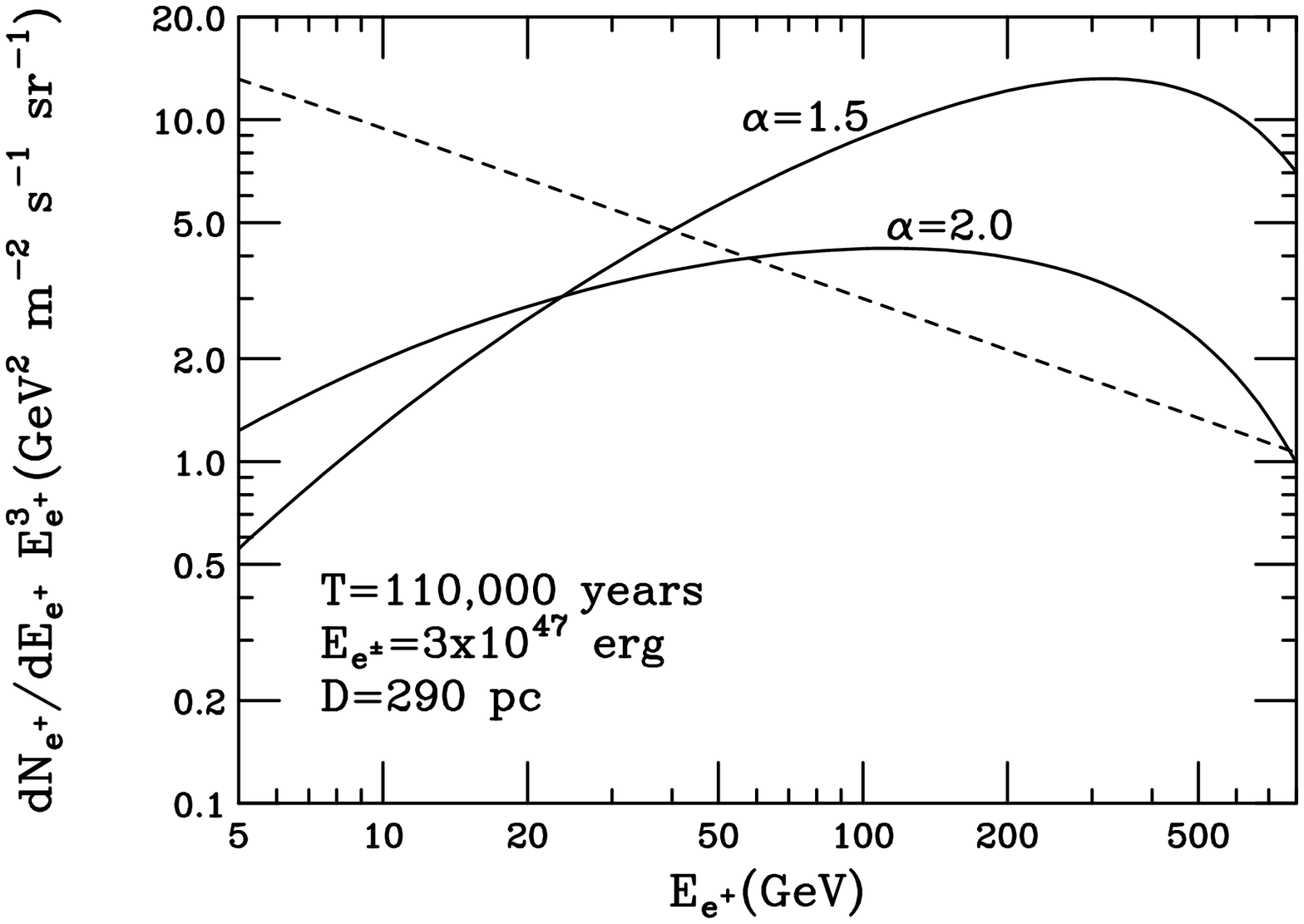}
\hspace{0.1cm}
\includegraphics[width=3.4in,angle=0]{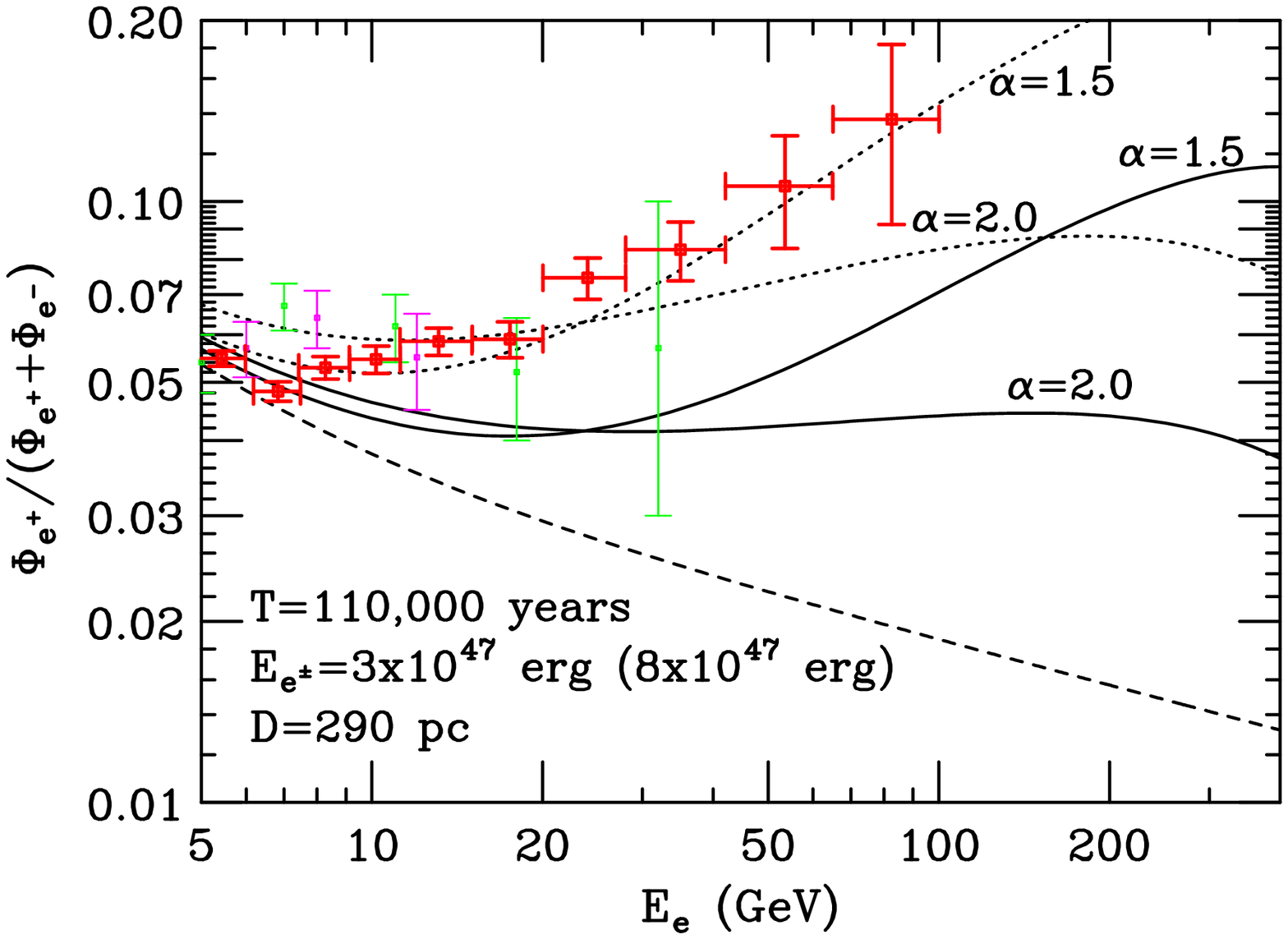}
\caption{As in Fig.~\ref{geminga}, but from the nearby pulsar
  B0656+14. The solid lines correspond to an energy in pairs given by
  $3\times 10^{47}$ erg, while the dotted lines require an output of
  $8\times 10^{47}$ erg.}  
\label{B0656+14}
\end{figure}

We now turn our attention to the case of the pulsar B0656+14. B0656+14
is considerably younger than Geminga (approximately $110,000$ years
old), has a period today of $P=390$ ms, and a current spin down
luminosity that is approximately the same as Geminga. The spectrum of
positrons and the positron fraction from B0656+14 are shown in the
left and right panels of Fig.~\ref{B0656+14}, respectively. The lines
are labeled as in Fig. \ref{geminga}. Because of the younger age of
this pulsar, the  
flux of positrons expected from B0656+14 is somewhat higher than from
Geminga, despite being somewhat more distant ($D=290$ pc). In the case
with $\alpha=1.5$, the predicted positron fraction can fit the PAMELA
data if it injected $\sim 8 \times 10^{47}$ erg in electron-positron
pairs. This appears still large, but less extreme than in the Geminga
case. 

\begin{figure}[t]
\centering\leavevmode
\includegraphics[width=3.4in,angle=0]{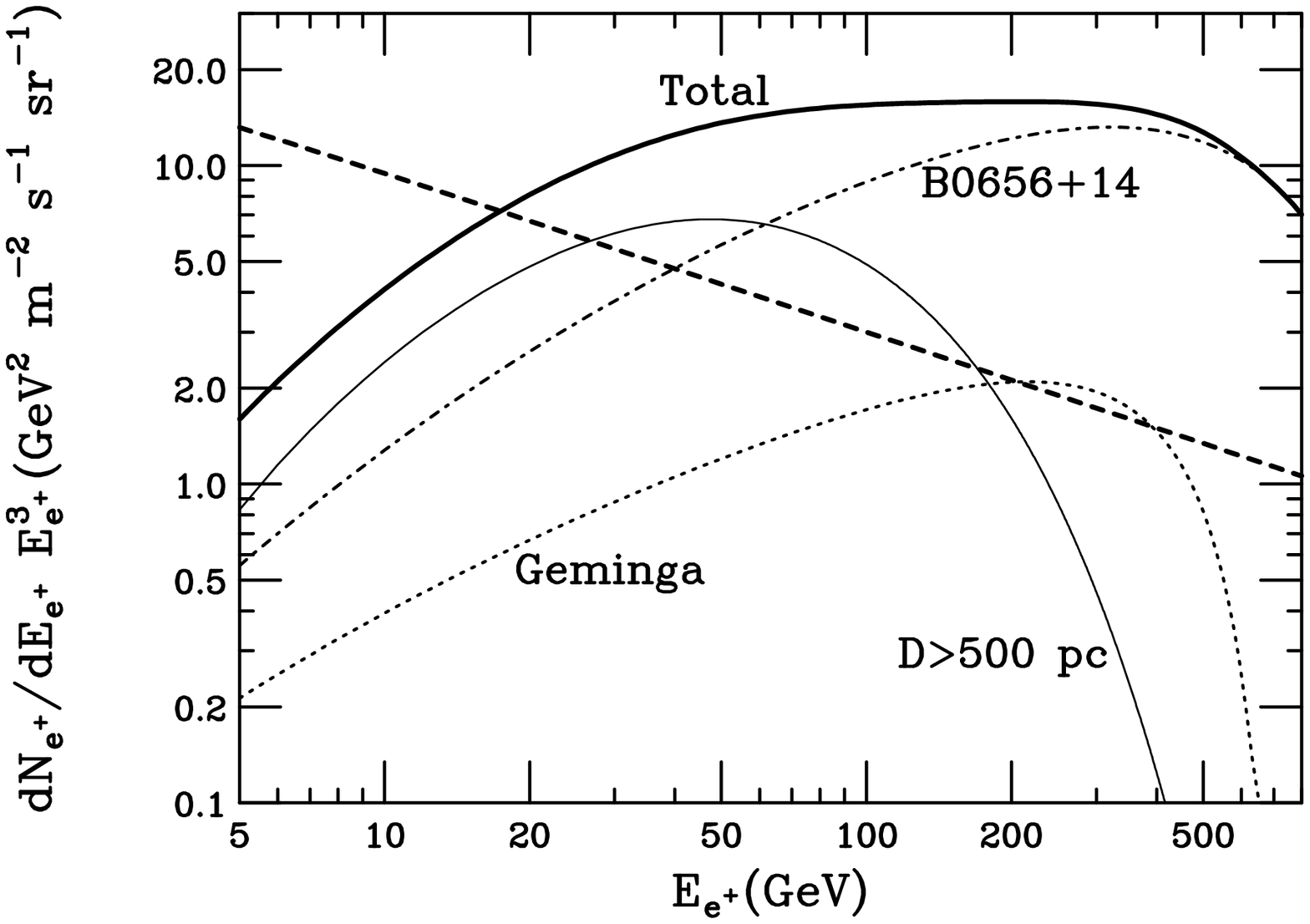}
\hspace{0.1cm}
\includegraphics[width=3.4in,angle=0]{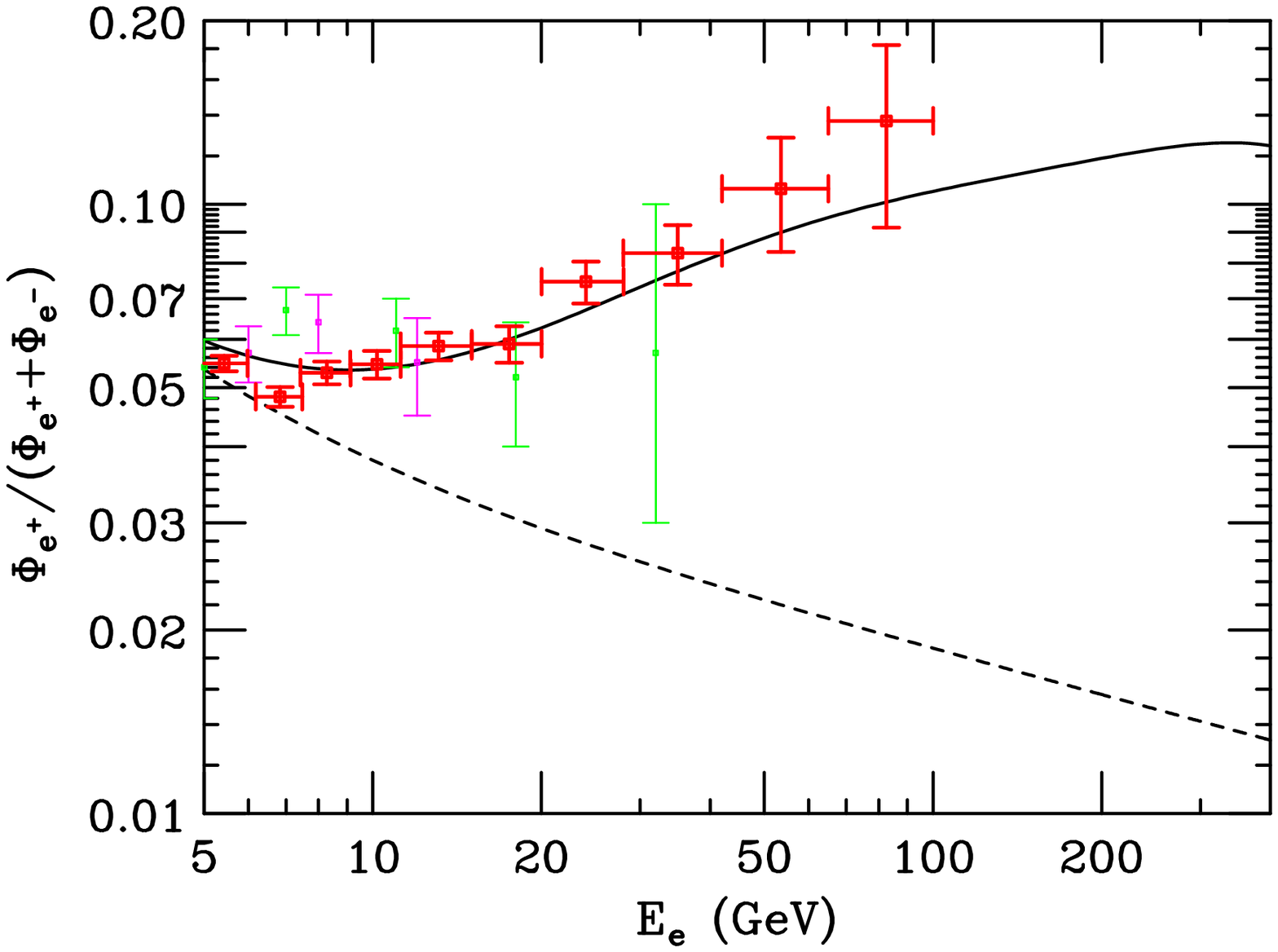}
\caption{The positron spectrum and positron fraction from the sum of
  contributions from B0656+14, Geminga, and all pulsars farther than
  500 parsecs from the Solar System.}  
\label{all}
\end{figure}

In Sec.~\ref{astrosource}, we found that nearby pulsars ($D\lsim500$
parsecs) are likely to dominate the pulsar contribution to the
positron spectrum, especially energies above $\sim50$ GeV. More
distant pulsars, however, are still anticipated to play an important
role in the lower energy range of the PAMELA positive excess. In
Fig.~\ref{all}, we show a combination of pulsar contributions to the
high energy positron spectrum and the positron fraction. In
particular, we include the contribution from all pulsars more distant
than 500 parsecs (using a rate of 4 pulsars per century, as shown in
the lower frames of Fig.~\ref{fig:sum}) and the contributions from
B0656+14 and Geminga (using $3\times 10^{47}$ erg in electron-positron
pairs from each and a spectral index of 1.5). As can be seen in the
right frame, such a combination can provide a good fit to the
preliminary measurements of PAMELA and accommodates for a rising
positron fraction even beyond $\sim 100\,$GeV.  

A comment regarding this result is in order. In
describing the contribution from all pulsars throughout the Milky Way,
we have adopted a spectrum with an index of 1.6 and an exponential
cutoff above 80 GeV. In contrast, we have used a higher cutoff (600
GeV) and slightly harder slope (1.5) for the nearby B0656+14 and
Geminga pulsars. These results do not contradict each other,  for the 
average injected spectrum from Ref.~\cite{zhangcheng}
results from the sum over a realization with wide variability of 
pulsar properties, including injected energy and spectral index.
In this respect, the results presented here are only demonstrative
of the fact that a significant role of nearby pulsars, while not {\it
  required} to explain present data, is consistent with them,  in
which case they should dominate the high energy tail. If this is the
case,  interesting observational 
signatures are possible, one of which is discussed in the next section.
In no case the high-energy spectra presented here should be considered as a 
robust prediction, since they depend crucially on the detailed
spectral properties 
of B0656+14, Geminga or and other nearby, mature pulsars that
contribute significantly to the high energy positron spectrum. 

\section{Distinguishing between pulsar and dark matter origins of high energy cosmic ray positrons}
\label{anisotropy}

The positron fraction reported by PAMELA taken alone is likely
insufficient to distinguish between dark matter and pulsar origins of
this signal. In this section we discuss an additional measurement
which may help to resolve this issue. In particular, even after the
diffusive propagation of electrons and positrons from pulsars is taken
into account, at sufficiently high energies a small dipole anisotropy
should be  
present in the direction of the dominant nearby source(s). In a very
general way, the anisotropy associated 
with diffusive propagation can be written as:  
\begin{equation}
\delta = \frac{I_{\rm max} - I_{\rm min}}{I_{\rm max} + I_{\rm min}} =
\frac{3 K |\nabla(dN_e/dE_e)|}{c\, (dN_e/dE_e)},
\end{equation}
where $\nabla (dN_e/dE_e)$ is the gradient of the electron/positron
density. The measurement of such an anisotropy in a statistical
significant manner requires a large number of electron/positron
events. For example, in order to detect an anisotropy at the $2\sigma$ level,
one needs to fulfill the condition $\delta \gsim 2\sqrt{2} (\dot
N_{\rm ev} t_{\rm obs})^{-1/2}$, where $\dot N_{\rm ev}$ is the rate of events
detected per unit time above a given threshold and $t_{\rm obs}$ is the
observation time. 

In addition to studying the gamma-ray sky, the Fermi gamma-ray
telescope will also be able to measure a flux of electrons (and
positrons, though without charge discrimination) at a rate of
approximately $3 \times 10^7$ electrons per year above 10
GeV~\cite{glastelectrons}. This implies that Fermi should be able to
detect (at the 2$\sigma$ confidence level) a dipole anisotropy in the
electron flux above 10 GeV if $\delta\gsim 0.05\%$ in one year or
$\delta\gsim 0.03\%$ in 5 years\footnote{Note that, extrapolating from the data
reported in~\cite{Adriani:2008zr}, PAMELA can collect $\sim 10^5$ useful 
electrons plus positrons events per year above 1.5 GeV. For the range
of interest 
here this is several orders of magnitude below Fermi and unlikely to lead to a 
meaningful constraint on the anisotropy of high energy charged leptons.}.   

\begin{figure}[t]
\centering\leavevmode
\includegraphics[width=3.5in,angle=0]{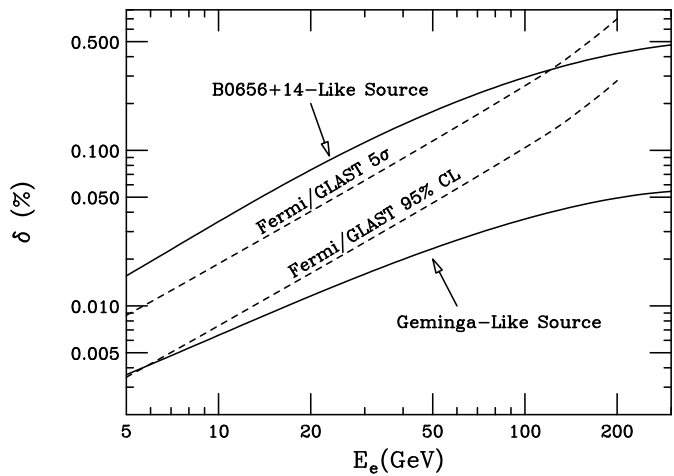}
\caption{The dipole anisotropy in the electron+positron spectrum from
  a source 110,000 years old at a distance of 290 pc (B0656+14-like)
  and from a source 370,000 years old at a distance of 157 pc
  (Geminga-like). In each case,  we have normalized the energy output
  to match the PAMELA data and have used a spectral shape of
  $dN_e/dE_e \propto E_e^{-1.5} \exp(-E_e/600\,\rm{GeV})$. Also shown
  as dashed lines is the sensitivity of the Fermi gamma-ray space
  telescope to such an anisotropy 
  (after five years of observation). The Fermi sensitivity shown is
  for the spectrum integrated above a given energy.} 
\label{anis}
\end{figure}

In Fig.~\ref{anis} we plot the level of anisotropy expected for a
Geminga-like and a B0656+14-like pulsar if they are responsible for
the majority of the observed positron excess. The two dashed lines
show the sensitivities of Fermi to anisotropy at 95\% confidence level
and at $5\sigma$ confidence level, after five years of observation
(integrated above a given energy). We find that Fermi should be
capable of identifying 
a single local source (or multiple sources in the same direction of
the sky) if that source injected the bulk of its electrons/positrons
within the last few hundred thousand years (the B0656+14-like and
Geminga-like cases correspond to injection 110,000 and 370,000 years
ago, respectively). If only a fraction of the high energy positrons
observed by PAMELA originate from a given nearby pulsar, the
corresponding solid lines shown in Fig.~\ref{anis} should be
multiplied (reduced) by this factor. Also note that B0656+14 and
Geminga lie in similar directions in the sky, so they are expected to
contribute the same overall dipole anisotropy.

Alternatively, if dark matter annihilations throughout the Milky Way's
halo are primarily responsible for the excess in the high energy
cosmic ray positron spectrum, a small dipole anisotropy in the
direction of the Galactic Center could also be generated. Fortunately,
both B0656+14 and Geminga are in approximately the opposite direction,
allowing for a potentially unambiguous discrimination between these
possibilities. In the special and relatively unlikely case that a
nearby dark matter subhalo in the direction of B0656+14/Geminga is
responsible for the observed flux, it would be difficult to
distinguish between pulsar and dark matter origins using this
technique. 

If anisotropy studies should prove inconclusive in resolving this
issue, other information could be inferred from the shape of the
positron fraction and of the overall electron/positron
spectrum. Peculiar shapes can result from the superposition of the
overall pulsar spectrum plus local contributions (see, for example,
Fig. 4 or Ref.~\cite{aharonian}). Future studies of the electron and
positron spectra at higher energies will be especially important, as
the spectral cutoff in the pulsar case is typically expected to be
smoother and less sudden than that predicted from annihilating dark
matter. Furthermore, combining electron/positron measurements with
those of antiprotons, antideuterons and diffuse gamma-rays may prove
useful in distinguishing between these possibilities. Population
studies of pulsars in the gamma-ray band by Fermi are also expected to
refine theoretical predictions and shed light on this issue.

\section{Conclusions}
\label{conclusions}

The results recently reported by PAMELA strongly indicate
the existence of a primary source or sources of high energy cosmic ray
positrons. This result is unexpected and very interesting, even if of
purely astrophysical origin. Several papers have appeared recently
which discuss this signal within the context of dark matter 
annihilations. In this article, we have instead explored the
possibility that the observed flux of high energy positrons is the
result of electron-positron pairs being produced in nearby and
galactic pulsars. We find that pulsars throughout the Milky Way, and a
small number of nearby mature pulsars, such as B0656+14 and Geminga,
could each plausibly generate the observed flux of positrons. To
normalize the overall flux, on the order of a few percent of the
pulsars' spin down power is required to be transferred into the
production of electron-positron pairs. The prediction in the case of
the sum of all pulsars in the Galaxy appears somewhat more robust in
that 
it relies on the average statistical properties of these astrophysical
objects rather than on the specific characteristics of nearby
pulsars.  It is remarkable that in this case, reasonable values for
the parameters can lead to a positron spectrum consistent with the
observations. Also, 
a pulsar origin would naturally fit the absence of an excess in the
anti-proton data~\cite{Adriani:2008zq}, 
since differently from dark matter scenarios no hadronic cascades are
associated with the production 
of pairs in the magnetospheres.

If a single or small number of nearby pulsars dominate the high energy
positron spectrum, an observable dipole anisotropy in the
electron-positron spectrum may be present. In particular, we find that
the Fermi gamma-ray space telescope would be able to detect the
anisotropy generated by a few hundred thousand year old or younger
source with greater than 2$\sigma$ significance. A $\sim$100,000 year
old source, such as B0656+14, would produce an anisotropy that could
be detected by Fermi with greater than 5$\sigma$ significance. This
provides a valuable test for 
distinguishing between pulsar and dark matter origins of the observed
cosmic ray positron spectrum, complementary to the spectral shape
and additional information from antiprotons and gamma-rays.

\acknowledgements{DH is supported by the Fermi Research Alliance, LLC
  under Contract No.~DE-AC02-07CH11359 with the US Department of
  Energy and by NASA grant NNX08AH34G.}

\end{document}